\documentclass
[superscriptaddress,secnumarabic,amssymb,amsmath,nobibnotes,aps,prd,showpacs,nofootinbib]{revtex4}%
\usepackage{graphicx,subcaption}
\usepackage{dcolumn}
\usepackage{bm}
\usepackage{xcolor}
\usepackage{amsmath}
\usepackage{amsfonts}
\usepackage{amssymb}
\usepackage{comment}
\usepackage{enumitem}
\usepackage{float}
\usepackage{hyperref}
\usepackage{footnote}
\usepackage{caption}

\newcommand{\ed}{\end{document}}
\newcommand{\beq}{\begin{equation}}
\newcommand{\eeq}{\end{equation}}
\newcommand{\beqa}{\begin{eqnarray}}
\newcommand{\eeqa}{\end{eqnarray}}
\newcommand{\bc}{\begin{center}}
\newcommand{\ec}{\end{center}}

\newcommand{\ba}{\begin{array}}
\newcommand{\ea}{\end{array}}

\begin{document}

\title{New Uncertainty Principle for a particle on a Torus Knot}

\author{Madhushri Roy Chowdhury}
\email{madhushrirc@gmail.com}
\affiliation{Physics and Applied Mathematics Unit, Indian Statistical Institute, 203 B. T. Road, Kolkata 700108, India}

\author{Subir Ghosh}
\email{subirghosh20@gmail.com}
\affiliation{Physics and Applied Mathematics Unit, Indian Statistical Institute, 203 B. T. Road, Kolkata 700108, India}

\begin{abstract}
 The present work deals with quantum Uncertainty Relations (UR) subjected to the Standard Deviations (SD) of the relevant dynamical variables. It is important to note that these variables have to obey the two distinct periodicities of the knotted paths embedded on the torus. We compute generalized forms of the SDs and the subsequent URs (following the Kennard-Robertson formalism). These quantities explicitly involve the torus parameters and the knot parameters where restrictions on the latter have to be taken into account. These induce restrictions on the possible form of wave functions that are used to calculate the SDs and URs and in our simple example, two distinct SDs and URs are possible. In a certain limit (thin torus limit), our results will reduce to the results for a particle moving in a circle.

 An interesting fact emerges that in the case of the SDs and URs, the local geometry of the knots plays the decisive role and not their topological properties.
\end{abstract}

\maketitle

\section{Introduction}
In recent times, the constrained motion of quantum particles on specified space curves embedded on specific surfaces, (see for example \cite{r1}), has developed into an active area of research. This is principally due to the extensive improvement in constructing geometric shaped channels from newly discovered materials, such as  carbon nanorings made from carbon nanotubes that are intimately connected to the edges of nanotubes, nanochains, graphene, and nanocones \cite{r4} having a variety of notable features such as multi-resonant properties, magneto-optical activity, paramagnetism, and ferromagnetism \cite{r5}. Practical applications of these nanostructures are in optical communications, isolators, traps for ions and atoms, and lubricants \cite{r6}. In the present work we will focus on quantum kinematics on a specific path - a torus knot   \cite{r7,r8} - that is, loosely speaking a one dimensional closed path (with certain well defined properties), embedded on the surface of a torus \cite{r2}. The  motivation for choosing this specific path of a torus knot comes from various urgent and topical issues, both theoretical as well as practical, that we briefly elaborate below. \\  

{\it{Aharonov-Bohm effect}}:  Long chain molecules, such as  polymers or oligomers (molecule that consists of a few repeating units which could be derived from smaller molecules, monomers), can form  entangled loops and knots causing an interaction between  molecular magnetism and conduction, leading to  the quantum-mechanical Aharonov-Bohm effect, that can change the electronic energy levels. Similar effect can be observed in  Knotted or entangled mesoscopic systems \cite{ab1,ab2,ab3}.\\
{\it{Optical activity}}: A recent exciting area of research is the optical activity, induced by knotted structures in molecules, the simplest being a trefoil knot, (studied in the present work as well), \cite{op1}. The paper  \cite{op2} reports on the creation of non-trivial knotted quantum wavefunctions of a quantum degenerate spinor Bose gas. On the other hand, in many optical and fluid studies, knots are formed as real space objects generating singular lines \cite{real}. the former example is analogous to condensed matter and field theory models, where the knots are formed in the mappings between the order-parameter space and real space \cite{ord} and is applicable to more general optical phenomena in atomic systems \cite{op3}.\\
{\it{Atomic or molecular realization}}: Theory of knots, and more broadly topological principles, have become an integral part of modern day physics, be it 
 from microscopic quantum systems to cosmology and elementary particle physics. In fact, probably the earliest impact of this theme was provided by Lord Kelvin, when he proposed the idea of  vortex atom theory  where linked vortex strings in the aether formed the structure of atoms \cite{mol1}. However, rightly so, this model of atoms was abandoned. Topological  implications  play major roles in  biophysical systems ranging from proteins to umbilical cords   \cite{umb} such that in some systems, there are mechanisms for a change in topology, resulting in a sudden change in physical properties \cite{umb1}. Molecular knots are found in nature mainly in DNA \cite{mol2} and protein molecules \cite{mol3}. Even though proteins and DNA display a rich variety of topological effects, both structural and dynamic, the function of and evolutionary advantage behind natural knots in biological systems are still strongly debated \cite{umb2}. \\
{\it{Quantum computation and information storing}}: The paper \cite{qcom} explores the interactions between knot theory and quantum computing. On one side, knot theory has been used to create models of quantum computing, and on the other, it is a source of computational problems. Several applications with optical-vortex knots have been implemented at macroscopic scales, such as the entanglement between topological features in a vortex link with single photons [4] and information transmission and storage \cite{qqcom}. Hence, these and other applications can be miniaturized to the fundamental optical-wavelength scale  provided that the knotted vortex fields reach those dimensions. Wavelength sized optical-vortex knots \cite{op4} could be even more important for future experiments or applications that require interaction between these fields and a few atoms [18].

The quantum mechanics of a particle on a torus knot was analyzed in \cite{r7}. Further, topological aspects such as the Berry phase and Hannay angle feature to this model were investigated in \cite{r8}; in particular, curvature and torsion effects of the knotted path were also taken into account.  

Uncertainty Relations (UR) play a key role both in understanding and applications of quantum mechanics. In the present work, we will shed some light on the UR imposed on the quantum mechanics of particles moving on a torus knot. At the outset, it should be emphasized that the URs we are going to study are, (in a sense), generalized forms of conventional URs because conventionally one considers URs between generic hermitian conjugate variables $A,B$, obeying the commutation relation $[A,B]=i\hbar \mu$ where $\mu$ can be a numerical constant. However, as we have discussed below briefly, (for details see Ogawa and Nagasawa \cite{on} that we have closely followed), for quantum systems in a closed configuration space, such as particle moving on a circle \cite{on} or on a torus knot as considered here, the naturally conjugate variables $$[\phi,L_z]=[\phi, -i\hbar \partial/\partial \phi =i\hbar$$ are not viable as the variable  $\phi$ does not obey the periodicity condition. This, in turn,  clashes with the hermiticity property of $L_z$ that is hermitian only in a Hilbert space consisting of periodic functions. The best one can do, (as suggested in \cite{on}), as one option, is to consider specific functions of $\phi$ that are explicitly periodic but not strictly conjugate to $L_z$ and burdens the RHS of $[\phi,L_z]$ with cumbersome factors of $\phi$. The alternate option \cite{on} is to work with Cartesian variables $X=Rcos\phi, Y=Rsin\phi$, that are explicitly periodic functions of $\phi$. In the following work, this prescription has been followed.

It has long been appreciated in quantum mechanics that sufficient care should be taken in deriving Uncertainty Relations (UR) for periodic degrees of freedom, the prime example being the UR for $\sigma_{\phi}\sigma_{L_{z}}$ where $\sigma_{\phi}$ and $\sigma_{L_{z}}$ stand for Standard Deviations (SD) for angle $\phi$ and its conjugate momentum $L_{z}$ respectively. For continuous variables ranging from $-\infty \rightarrow +\infty$, such as position $x$ and momentum $p_x = -i\hbar \partial/\partial x$, satisfying $[x,p_x]=i\hbar$, using conventional definitions for SD for a generic variable $A,~ \sigma_{A} = \sqrt{\langle  (A- \langle A \rangle)^{2} \rangle} = \sqrt{\langle A^{2}\rangle - \langle A \rangle^{2}}$, (with $\langle A \rangle =  \int _{-\infty}^{\infty} \psi(x)^* A \psi(x)dx$), one derives 
\begin{equation}
\sigma_{x}\sigma_{p_{x}} \geq \hbar /2 .
\label{eqn1}
\end{equation}
Clearly a naive replacement of $x,p_x$ by $\phi, L_Z$, obeying $\left[ \phi, \hat{L}_{z}\right] =i \hbar$ will lead to 
\begin{equation}
\sigma_{\phi}\sigma_{L_{z}} \geq \hbar /2
\label{eqna1}
\end{equation}
which, however, is inconsistent since, for angular momentum eigen functions $\sigma_{L_{z}}=0$ but the angle $\phi$, being periodic, is restricted to $2\pi\geq \phi \geq 0$. The problem lies with the treatment of $\phi$ which is continuous and in principle can run from $-\infty $ to $+\infty $ but at the same time physics has to be the same for $\phi =\beta$ and $\phi =\beta + 2n\pi, n$ an integer. Various inequivalent solutions have been proposed \cite{r9}. Recent works in related areas are \cite{r10}. 

In the present paper, we will closely follow the framework proposed by Ogawa and Nagasawa \cite{on} that suggests working in terms of variables that are {\it{explicitly periodic functions of }} $\phi$. In \cite{on} the problem of a quantum particle on a circle and its corresponding URs are studied in detail. In the present work,  a non-trivial extension of the above, a quantum particle on a Torus knot, will be considered. It is important to notice that although the particle on a circle problem can be stated with one degree of freedom, it requires the Cartesian $X-Y$-plane with a two-variable parameterisation to correctly account for the restricted domain of $\phi$ \cite{on}. In the same spirit, in our model, although the particle on a torus knot path embedded on a torus can be described by one variable $\phi$ (due to the torus knot constraint), it requires the Cartesian $X-Y-Z$ space with a three-variable parameterisation to correctly account for the restricted domain of $\phi$ \cite{r7,r8}. Furthermore, the novelty also comes from the explicit presence of the torus knot parameters in the URs that can influence the latter. These works pertain to a general remark by Ogawa  that "The uncertainty broadens out the space!" \cite{oga} since due to quantum correlations, the mean position of the particle is at a coordinate that is not in the particle's configuration space (similar to the centre of mass of a ring being at the centre, outside the ring). 
\section{Particle on $S_1$ (circle)}
Let us make a quick digression to the problem of particle on $S_1$ \cite{on}  before moving on to our more complex problem at hand,  particle on a torus knot; essentially same framework will be exploited in the latter problem.  Inherently periodic Cartesian coordinates $  X=Acos\phi, Y=Asin\phi $ are used as basic position variables instead of $ \phi $. Since $X^2+Y^2=A^2$ and $\langle X\rangle^{2} \leq \langle X^{2}\rangle ,~ \langle Y\rangle^{2} \leq \langle Y^{2}\rangle $, the authors of \cite{on} define the "Mean Resultant Length" (MRL) $\mathcal{A}$
\begin{equation}
\mathcal{A}= \sqrt{\langle X\rangle^{2}+ \langle Y\rangle^{2}}
\end{equation}
agreeing to the inequality 
\begin{equation}
\mathcal{A}= \sqrt{\langle X\rangle^{2}+ \langle Y\rangle^{2}} \leq \sqrt{\langle X^{2}+ Y^{2}\rangle}=A .
\end{equation}
Following the definition $\langle \{X,Y\} \rangle =  \int _{0}^{\infty}\{X,Y\} |\psi(\phi)|^2 d\phi$ \cite{on}, let us introduce
 $ \langle \phi \rangle $ when $ \mathcal{A} \neq 0 $, as $
\langle X\rangle = \mathcal{A}~ cos \langle \phi\rangle ,~
\langle Y\rangle = \mathcal{A}~ sin \langle \phi\rangle $.
Hence for uniform amplitude $ \mid \psi \mid $ one has $\mathcal{A}=0$ indicating that the mean position of the particle is at the origin of the circle. Note that, for the circular configuration space of the particle, all positions on the circle are equivalent and hence the mean position, represented by a specific angle is not acceptable. Treating the centre of the circle as the mean position can be justified through an analogy of the centre of mass of a uniform wire: in the form of a straightline, the centre of mass is the middle of the wire or "zero" position that is within the configuration space of the wire, whereas in the form of a circle, the centre of mass is at the centre of the wire, {\it{outside}} the configuration space \cite{on}.  
 
To construct URs involving the SDs,  using  
 $ \Delta X= X- \langle X \rangle, ~ \Delta Y= Y- \langle Y \rangle,~ \Delta \hat{L}_{z}= \hat{L}_{z}- \langle {L}_{z} \rangle $, one has
  \begin{equation}
  \sigma^{2}_{X}= \langle (\Delta X)^{2}\rangle = \langle X^{2}\rangle- \langle X\rangle^{2}, ~
  \sigma^{2}_{Y}= \langle (\Delta Y)^{2}\rangle = \langle Y^{2}\rangle- \langle Y\rangle^{2}
 \end{equation} 
 and  the commutation relations  
 \begin{equation}
 \left[ X, {L}_{z}\right]=-i \hbar Y,~
 \left[ Y, {L}_{z}\right]=i \hbar X .
 \label{eqn22}
 \end{equation}
To  follow the Kennard-Robertson \cite{kr} method (see also \cite{sch}), a real parameter $ \lambda $ is introduced such that
\begin{equation}
 \mid \mid (\Delta X - i\lambda \Delta {L}_{z})\mid \psi \rangle \mid \mid^{2} = \int \left[\lbrace (\Delta X- i \lambda \Delta {L}_{z})\psi \rbrace^{\ast} \lbrace (\Delta X- i \lambda \Delta {L}_{z})\psi\rbrace\right]dx
 \label{k1}
 \end{equation}
 yielding 
 \begin{equation}
      \langle \psi \mid (\Delta X + i \lambda \Delta {L}_{z})(\Delta X - i \lambda \Delta {L}_{z})\mid \psi \rangle \geq 0 
      \label{k2}
  \end{equation}
  leading to
 \begin{equation}
       \Rightarrow \langle \psi \mid (\Delta X)^{2} \mid \psi \rangle + \lambda^{2} \langle \psi \mid (\Delta {L}_{z})^{2} \mid \psi\rangle -i \lambda \langle \psi \mid [\Delta X, \Delta {L}_{z}] \mid \psi \rangle \geq 0 
 \end{equation}
 Using $ [\Delta X, \Delta {L}_{z}]=[X,{L}_{z}]=-i \hbar Y $, we get,
 \begin{equation}
   \langle \psi \mid (\Delta X)^{2} \mid \psi \rangle + \lambda^{2} \langle \psi \mid (\Delta {L}_{z})^{2} \mid \psi\rangle - \hbar \lambda \langle \psi \mid Y \mid \psi \rangle \geq 0 .  
 \end{equation}
   This inequality will hold for any real value of $ \lambda $, provided  the discriminant is $\leq  0$ so that
 \begin{equation}
 (n \hbar)^{2} \langle \psi \mid Y \mid \psi \rangle^{2} - 4\langle \psi \mid (\Delta X)^{2} \mid \psi \rangle\langle \psi \mid (\Delta {L}_{z})^{2} \mid \psi\rangle \leq 0 
 \label{k4}
 \end{equation}
 Finally, the URs are obtained as 
  \begin{equation}
 \sigma_{X}\sigma_{{L}_{z}} \geq \frac{ \hbar}{2}\mid\langle Y \rangle\mid ,~
  \sigma_{Y}\sigma_{{L}_{z}} \geq \frac{ \hbar}{2}\mid\langle X \rangle\mid
 \label{eqn26}
 \end{equation}
 
 This approach will be followed by us in deriving the torus knot URs.
\section{Torus, Torus Knot and Uncertainty Relations}
A torus is the surface of revolution generated by revolving a circle in 3D space about an axis coplanar with the circle. 
If the axis of revolution does not touch the circle, then the surface has a ring shape and it is called the torus of revolution. Torus knots are a special class of closed curves wound on the surface of a geometric torus in $ \Re^{3} $. 
In a $(p,q)$ torus knot, $p$ denotes the number of windings in the toroidal direction(about the axis of rotational symmetry of the torus) and $q$ denotes the number of windings in the poloidal directions (about the cross-sectional axis of symmetry of the torus). Both $p$ and $q$ are natural numbers and they are coprime meaning that the only positive integer that is a divisor of both $p,q$ is $1$. A torus knot is trivial iff either $p$ or $q$ is equal to $1$ or $-1$ \cite{r2,r7,r8}.

Let us define $ \alpha = -\frac{q}{p} $ to be the winding number of (p,q) torus knot, that connects the angles $\theta$ and $\phi$ through $p\theta + q\phi =0$, where $\eta, \theta,\phi$ constitute the toroidal coordinates, as given below. It is a measure of the complexity and unique identity of the (p,q) torus knot. The trefoil knot or the (2,3) torus knot is the simplest non-trivial knot (see Figure).

The toroidal coordinates are defined as,
\begin{equation}
x= \frac{a ~sinh \eta ~ cos \phi}{cosh \eta - cos \theta},~
y= \frac{a~ sinh \eta ~ sin \phi}{cosh \eta - cos \theta},~
z= \frac{a~ sin \theta}{cosh \eta - cos \theta}
\end{equation}
where $ 0< \eta< \infty $, $ -\pi < \theta < \pi $ and $ 0 < \phi < 2 \pi $.

Now, $ \eta = \eta_{0} $ fixes the toroidal surface on which the knot is wound and by imposing the winding number condition $ \theta = \alpha \phi $, we have only one independent variable $ \phi $. Thus a particular $\alpha$ torus knot is characterized by 
\begin{equation}
x= \frac{a~ sinh \eta_{0}~ cos \phi}{cosh \eta_{0} - cos (\alpha \phi)},~
y= \frac{a ~sinh \eta_{0}~ sin \phi}{cosh \eta_{0} - cos (\alpha \phi)}~
z= \frac{a~ sin (\alpha \phi)}{cosh \eta_{0} - cos (\alpha \phi)}
\label{c0}
\end{equation}
where $ 0< \phi < 2 \pi p $.
The major radius and the minor radius of a torus are represented by R and d respectively. We define the parameter $ a^{2} = R^{2} - d^{2} $ and the aspect ratio to be $R/d= cosh (\eta_{0}) $.

The required commutation relations will be
\begin{equation}
[x, {L}_{z}]= i \hbar \frac{\partial x}{\partial \phi},~~
[y, {L}_{z}]= i \hbar \frac{\partial y}{\partial \phi},~~
[z, {L}_{z}]= i \hbar \frac{\partial z}{\partial \phi}.
\label{cx1}
\end{equation}
It is easy to see that using the relations (\ref{c0}), the above commutation relations will be very complicated.  First, we use  a compact notation
\begin{equation}
x= \frac{a \beta cos \phi}{\gamma - cos (\alpha \phi)},~~ 
y= \frac{a \beta sin \phi}{\gamma - cos (\alpha \phi)},~~
z= \frac{a sin (\alpha \phi)}{\gamma - cos (\alpha \phi)}
    \label{3}
\end{equation}
where $\beta =sinh \eta_{0}, \gamma = cosh \eta_{0}$ and 
\begin{equation}
x^2+y^2+z^2=a^2\frac{\beta^2+sin^2(\alpha\phi)}{\left(\gamma -cos(\alpha\phi)\right)^2}.
    \label{a3}
\end{equation}
 Since $q,p$ are integers but $\alpha =-q/p$ is not necessarily an integer, $x,y,z$ are periodic under $\phi \rightarrow \phi + 2p\pi$. In the present work, we will restrict ourselves to positive $p,q$ values. 
\begin{figure}[h]
\begin{center}
\includegraphics[scale =0.5]{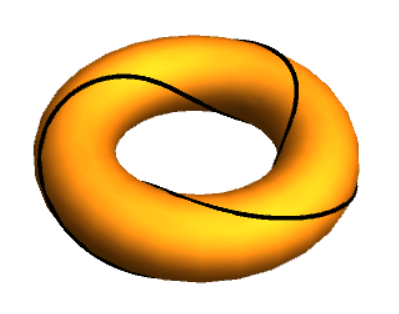}
\caption*{Figure: The $(2,3)$ Trefoil knot}
\end{center}
\end{figure}

 To produce analytical results and for a comparison with the results obtained for particle on circle \cite{on}, we invoke the popular thin torus approximation with $a>>0$ and a large aspect ratio, i.e.,  large $\eta_0$ so that to $O(1/\eta_0 )$, $\beta ={\sqrt {\gamma^2-1}}\approx \gamma -1/\gamma$. Thus we now work with the parameterisations, 
\begin{equation}
x=a~cos\phi \left(1 +\frac{cos(\alpha\phi)}{\gamma}\right), ~~
y=a~sin\phi \left(1 +\frac{cos(\alpha\phi)}{\gamma}\right), ~~
z=a~\frac{sin(\alpha\phi)}{\gamma}, 
    \label{5}
\end{equation}
with $L_z=-i\hbar\partial/\partial \phi$.By dropping the $1/\gamma$-terms, a circle of radius $a$ is recovered.  Note that the same periodicity $\phi \rightarrow \phi +2p\pi$ is maintained. The relation (\ref{a3}) reduces to
\begin{equation}
x^2+y^2+z^2=a^2(1+\frac{2}{\gamma}cos(\alpha\phi))
    \label{6}
\end{equation}
(with $(a/\gamma)^2\approx 0$), which can further be rewritten as
\begin{equation}
x^2+y^2+z^2=a^2\left(1+\frac{2}{\gamma}{\sqrt{1-\left(\frac{\gamma z}{a}\right)^2}}~\right).
    \label{x7}
\end{equation}
Notice that for coordinates on the torus knot, in the thin torus approximation, in (\ref{5}), $z<<x,y$ since $z$ has a factor of $1/\gamma$. Finally  (\ref{x7}) simplifies to
\begin{equation}
x^2+y^2+(1+\gamma )z^2=a^2\left(1+\frac{2}{\gamma}\right).
    \label{8}
\end{equation}
Using  (\ref{cx1}) in the Kennard-Robertson approach, the commutators in this  approximation turn out to be
\begin{equation}
[x,L_z]=-i\hbar\left(y+\frac{\alpha }{a}zx\right),~
[y,L_z]=i\hbar\left(x-\frac{\alpha }{a}zy\right),~[z,L_z]=i\hbar\alpha \left(\frac{a}{\gamma}-\frac{\gamma}{2a}z^2\right).
    \label{9}
\end{equation}
The cherished forms of URs for the particle on a $\alpha$-torus knot are given by
\begin{equation}
\sigma_x^2\sigma_{L_z}^2\geq \frac{\hbar ^2}{4}\left<y+\frac{\alpha }{a}zx\right>^2,~
\sigma_y^2\sigma_{L_z}^2\geq  \frac{\hbar ^2}{4}\left<x-\frac{\alpha }{a}zy\right>^2,~
\sigma_z^2\sigma_{L_z}^2\geq  \frac{\hbar ^2\alpha^2}{4}\left<\frac{a}{\gamma}-\frac{\gamma}{2a}z^2\right>^2 .
    \label{10}
\end{equation}
It is worthwhile to note that for $\alpha =0$ the torus knot disappears and as expected, our results reduce to the simpler results of a particle on a circle (\ref{eqn26}) \cite{on}. Another significant observation is that the URs depend on the local geometric structure of the knots and not on their topological properties. It is well known that the $(p,q)$ knot is topologically equivalent to the $(q,p)$ knot, which is not reflected in (\ref{10}) since the results will change under $\alpha \rightarrow 1/\alpha $. 

These results are new and constitute an important part of our work.
\subsection{Application: Standard Deviations }
 $L_z$-eigenfunctions $\psi_n$ will be of the form
\begin{equation}
\psi_n=\frac{1}{{\sqrt{2p\pi}}}e^{i\frac{n}{p}\phi}
    \label{x2}
\end{equation}
for integer $n$. The $p$-dependence results from the specific periodicity connected to the $p,q$ torus knot. Let us consider a superposition $\psi$ of two such solutions
\begin{equation}
\psi=\frac{1}{{2\sqrt{p \pi}}}(e^{i\frac{n}{p}\phi}+e^{i\frac{k}{p}\phi})
    \label{x3}
\end{equation}
and use it to compute the expectation values
\begin{equation}
<x>=\int_0^{2p\pi} \psi^*~ x ~ \psi ~d\phi ,~~<L_z>=\int_0^{2p\pi} \psi^*~ L_z ~ \psi ~ d\phi
    \label{x4}
\end{equation}
We recall that $ \alpha =-q/p$ and $ n,k,q,p $ all are integers and with
\begin{equation}
<x> = \int_0^{2 p \pi} \left(\frac{1}{{2\sqrt{p \pi}}}\right)^2(e^{-i\frac{n}{p}\phi}+e^{-i\frac{k}{p}\phi})\left(a~ cos \phi \left(1 + \frac{cos(\alpha \phi)}{\gamma}\right)\right)(e^{i\frac{n}{p}\phi}+e^{i\frac{k}{p}\phi}) d \phi
\label{xx}
\end{equation}
and similarly for $<y>, <z>$, the results are
\begin{equation}
<x>=  \frac{a}{2 \gamma}\delta^{1}_{\alpha}+ \frac{a}{2} \delta^{1}_{\frac{|k-n|}{p}}  + \frac{a}{4 \gamma} \left(\delta^{1}_{\alpha + \frac{|k-n|}{p}} + \delta^{1}_{\alpha - \frac{|k-n|}{p}}\right) , ~~  <y>= 0, ~~ 
<z>= 0,
    \label{sz}
\end{equation}
From the non-triviality condition of a torus knot, $p\neq 1, q\neq 1, p\neq q, \alpha \neq 1$, the first term in RHS of $<x>$ vanishes thereby leading to
\begin{equation}
<x>=  \frac{a}{2} \delta^{1}_{\frac{|k-n|}{p}}  + \frac{a}{4 \gamma}\left( \delta^{1}_{\alpha + \frac{|k-n|}{p}}+\delta^{1}_{\alpha - \frac{|k-n|}{p}}\right)  , ~~  <y>= 0, ~~ 
<z>= 0.
    \label{asz}
\end{equation}
Let us rewrite $<x>$ as
\begin{equation}
<x>=  \frac{a}{2} \delta^{p}_{|k-n|}  + \frac{a}{4 \gamma}\left( \delta^{(p+q)}_{|k-n|}+\delta^{(p+q)}_{-|k-n|}\right) , ~~  <y>= 0, ~~ 
<z>= 0.
    \label{absz}
\end{equation}
Furthermore, since $p,q$ are positive integers, we finally obtain
\begin{equation}
<x>=  \frac{a}{2} \delta^{p}_{|k-n|}  + \frac{a}{4 \gamma} \delta^{(p+q)}_{|k-n|} , ~~  <y>= 0, ~~ 
<z>= 0.
    \label{abcsz}
\end{equation}
Expectation value of $<L_z>$ turns out to be
 $$<L_z> = \int_0^{2 p \pi} \frac{1}{{2\sqrt{p \pi}}}(e^{-i\frac{n}{p}\phi}+e^{-i\frac{k}{p}\phi})\left(- i \hbar \frac{\partial}{\partial \phi}\right)\frac{1}{{2\sqrt{p \pi}}}(e^{i\frac{n}{p}\phi}+e^{i\frac{k}{p}\phi}) d \phi$$
\begin{equation}
= \frac{(n+k)}{2p}\hbar ,~~ k \neq n .
    \label{sl}
\end{equation}
It will be interesting to compare the above results with those obtained in \cite{on} for particle on a circle,
\begin{equation}
<X>=\frac{1}{2}\delta^1_{|n-k|},~~<Y>=0,~~<L_z>=\frac{(n+k)}{2}\hbar
\label{s2}
\end{equation}
(i) For the torus nearly reducing to a circle (extremely thin torus) $a\approx R,  1/\gamma \approx 0$ and $p=1$ (no non-trivial knot) our results (\ref{asz}) reduces to the circular case (\ref{s2}), as discussed in \cite{on}.\\
(ii) The two expressions for $L_z$ in (\ref{sl}) and (\ref{s2}) become identical.\\
(iii) For a knot, $p,q$ are integers ($\neq 1$), $<L_z>$ for particle on a knot (\ref{sl}) is smaller than $<L_z>$ for particle on a circle (\ref{s2}). For all $<X>=<Y>=0$ or $<x>=<y>=<z>=0$ the particle is effectively at the centre of the circle or torus respectively. To obtain a non-zero mean position, the superposition in $\psi$ has to follow restrictions: on the circle $|n-k|=1$ and two possibilities on the torus $|n-k|=p$ or  $|n-k|=p\pm q$. Notice  {\it{the knot parameters are explicitly entering the outcomes in a non-trivial way}}. Obviously numerical values will also differ due to the torus and circle parameters in the expressions.

In order to compute Standard Deviations (SD)
\begin{equation}
\sigma_x^2 =<x^2>-<x>^2,~~\sigma_y^2 =<y^2>-<y>^2, ~~\sigma_z^2 =<z^2>-<z>^2
\label{sd1}
\end{equation}
we first calculate 
\begin{equation}
<x^{2}> = \frac{a^{2}}{2} ,~<y^{2}> = \frac{a^{2}}{2} + \frac{a^{2}}{4 \gamma^{2}},~<z^{2}> = \frac{a^{2}}{2 \gamma^{2}} .
\label{sd2}
\end{equation}
It is somewhat surprising to find that the above results do not depend on the knot parameters (as these conspire to get cancelled), at least in the present level of approximation and the particular form of $\psi$ considered.

From the result
\begin{equation}
<L_z^{2}> = \frac{(n^{2}+k^{2})\hbar^{2}}{2 p^{2}}  
\label{sd5}
\end{equation}
we get 
\begin{equation} 
\sigma_{L_z}^2=\frac{(n^2+k^2)\hbar ^2}{2p^2}-\left(\frac{(n+k)}{2p}\hbar \right)^2=\frac{\hbar^2}{4p^2}(n-k)^2.
\label{sd6}
\end{equation}
Considering the two possible restrictions\\
\begin{equation}
  (I)~~ |n-k|=p: ~  \sigma_x\approx\frac{a}{2},~\sigma_y\approx\frac{a}{{\sqrt{2}}},~ \sigma_z=\frac{a}{\sqrt{{2}}\gamma},~\sigma_{L_z}=\frac{\hbar}{2}
  \label{sig}
\end{equation}
\begin{equation}
  (II)~~ |n-k|=p+q: ~  \sigma_x\approx\frac{a}{{\sqrt{2}}},~\sigma_y\approx\frac{a}{{\sqrt{2}}},~ \sigma_z=\frac{a}{\sqrt{{2}}\gamma},~\sigma_{L_z}=\frac{\hbar (p+q)}{2p}.
  \label{sig1}
\end{equation}
Interestingly, $\sigma_{L_z}$ for the choice (I) is the same as the circular case for $|n-k|=1$ \cite{on}. 

We exhibit the results obtained so far in a table,
\begin{table}
\begin{center}
\begin{tabular}{|c|c|c|}
\hline
expec. values & $(I): p=|n-k|$ & $(II): p+q=|n-k|$\\
\hline
$<x>$ & $\frac{a}{2}$ & $\frac{a}{4\gamma}$\\
\hline
$<y>$ & 0 & 0\\
\hline
$<z>$ & 0 & 0\\
\hline
$<x^2>$ & $\frac{a^2}{2}$ & $\frac{a^2}{2}$\\
\hline
$<y^2>$ & $\frac{a^2}{2}+\frac{a^2}{4\gamma^2}$ & $\frac{a^2}{2}+\frac{a^2}{4\gamma^2}$\\
\hline
$<z^2>$ &   $\frac{a^2}{2\gamma^2}$ &$\frac{a^2}{2\gamma^2}$\\
\hline
$<\sigma_x>$ & $\frac{a}{2}$ &$\frac{a}{{\sqrt{2}}}$ \\
\hline
$<\sigma_y>$ & $\frac{a}{{\sqrt{2}}}$ & $\frac{a}{{\sqrt{2}}}$\\
\hline
$<\sigma_z>$ & $\frac{a}{{\sqrt{2}\gamma}}$ & $\frac{a}{{\sqrt{2}\gamma}}$\\
\hline
$<L_z>$ & $\frac{\hbar(n+k)}{2p}$ & $\frac{\hbar(n+k)}{2p}$\\
\hline
$<L^2_z>$ & $\frac{(n^2+k^2)\hbar^2}{2p^2}$ & $\frac{(n^2+k^2)\hbar^2}{2p^2}$\\
\hline
$<\sigma_{L_z}>$ & $\frac{|n-k|\hbar}{2p}=\frac{\hbar}{2}$ & $\frac{(p+q)\hbar}{2p}$\\
\hline
\end{tabular}
\caption*{}{Table: Expectation values and Standard Deviations}
\label{Table 1.1}
\end{center}
\end{table}
\subsection{Application: Uncertainty Relations}
Let us return to the URs derived earlier in (\ref{10})
\begin{equation}
\sigma_x\sigma_{L_z}\geq \frac{\hbar }{2}\left<y+\frac{\alpha }{a}zx\right> = \frac{\hbar }{2}\left<y-\frac{q }{pa}zx\right>,
\label{ur1}
\end{equation}
\begin{equation}
\sigma_y\sigma_{L_z}\geq  \frac{\hbar }{2}\left<x-\frac{\alpha }{a}zy\right>=\frac{\hbar }{2}\left<x+\frac{q }{pa}zy\right>,
\label{ur2}
\end{equation}
\begin{equation}
\sigma_z\sigma_{L_z}\geq  \frac{\hbar \alpha }{2}\left<\frac{a}{\gamma}-\frac{\gamma}{2a}z^2\right>=-\frac{q}{p}\frac{\hbar }{2}\left<\frac{a}{\gamma}-\frac{\gamma}{2a}z^2\right> .
    \label{ur3}
\end{equation}
To explicitly  calculate the above, in the LHS we need the following: 
\begin{equation}
<zx> = 0, 
    \label{r1r}    
    \end{equation}
$$<zy> = \frac{a^{2}}{2 \gamma} \delta^{1}_{\alpha} + \frac{a^{2}}{4 \gamma^{2}} \delta^{1- \alpha}_{\alpha} + \frac{a^{2}}{4 \gamma} \left(\delta^{\alpha}_{1+ \frac{k-n}{p}} +  \delta^{\alpha}_{1-\frac{k-n}{p}}\right)    
$$
\begin{equation}
+ \frac{a^{2}}{8 \gamma^{2}}\left( \delta^{\alpha}_{1+\alpha+\frac{k-n}{p}}+ \delta^{\alpha}_{1+\alpha-\frac{k-n}{p}}
    +  \delta^{\alpha}_{1-\alpha+\frac{k-n}{p}} +  \delta^{\alpha}_{1-\alpha-\frac{k-n}{p}}\right).
    \label{rr}
    \end{equation}
Pertaining to the two choices, $<zy>$ in (\ref{rr}) reduces to
\begin{equation}
(I)~\rightarrow~|n-k|=p~, <zx>=0, <zy> = \frac{a^2}{8\gamma^2};~~(II)~\rightarrow~|n-k|=p+q~, <zx>=0, <zy> =\frac{ a^2}{4\gamma}.
    \label{r1r}    
    \end{equation}
Our aim is to compute the LHS and RHS independently for each of the URs (\ref{ur1},\ref{ur2}, \ref{ur3} ) and check their validity. \\
(\ref{ur1}):  Our first result is that the UR involving $\sigma_x\sigma_{L_z}$ is trivially satisfied since the RHS is zero (to the approximation we are working). \\
(\ref{ur2}): Analysis for the UR involving $\sigma_y\sigma_{L_z}$ for the two possible options:\\
Choice (I): LHS =$\frac{a\hbar}{2{\sqrt{2}}};~~RHS=\frac{a\hbar}{4}(1+\frac{q}{4p\gamma^2})\approx \frac{a\hbar}{4} $. Clearly, the inequality is satisfied for choice I.\\
Choice (II): LHS =$\frac{a \hbar}{2{\sqrt{2}}}(1+\frac{q}{p} );~~RHS=\frac{a\hbar}{8\gamma}(1+\frac{q}{p})$ \\
A simple algebra reveals that $LHS-RHS\approx \frac{a\hbar}{8}(1+\frac{q}{p})(2{\sqrt{2}}-\frac{1}{\gamma})$. Clearly the inequality is satisfied for choice II as well (since $1/\gamma$ is small).\\

(\ref{ur3}): Analysis for the UR involving $\sigma_z\sigma_{L_z}$ for the two possible options:\\
Choice (I): LHS =$\frac{a\hbar}{2{\sqrt{2}}\gamma};~~RHS=-\frac{3a\hbar q}{s\gamma p}~\rightarrow LHS-RHS=\frac{a\hbar}{8\gamma}(2{\sqrt{2}}+\frac{3q}{p})$. Clearly, the inequality is satisfied for choice I.\\
Choice (II): LHS =$\frac{a \hbar}{2{\sqrt{2}}\gamma}(1+\frac{q}{p} );~~RHS=-\frac{3a\hbar q}{8\gamma p}~\rightarrow LHS-RHS=\frac{a\hbar q}{8\gamma }\left(2{\sqrt{2}}(1+\frac{q}{p})+\frac{3q}{p}\right)$\\
Clearly, the inequality is satisfied for choice II as well.

To conclude this section, we note that we have derived new structures of Uncertainty Relations applicable to a quantum particle on a torus knot. These are non-trivial extensions of similar results for a particle on a circle, discussed in \cite{on}. We have considered a specific form of $\psi$ to explicitly show that all the URs are satisfied. These constitute part of our main results.
 \subsection{Application: Mean Resultant Length}  
Following \cite{on} let us define a Mean Resultant Length (MRL)
\begin{equation}
R\equiv {\sqrt{<x>^2 +<y>^2+(1+\gamma ) <z>^2}} \leq {\sqrt{1+\frac{2}{\gamma}}}~a\approx \left(1+\frac{1}{\gamma} \right)a
    \label{11}
\end{equation}
which holds because for each $j,~<x_j>^2<<x_j^2>$ and hence 
$$
R={\sqrt{<x>^2 +<y>^2+(1+\gamma ) <z>^2}}\leq {\sqrt{<x^2> +<y^2>+ (1+\gamma )<z^2>}} $$
\begin{equation}
={\sqrt{<x^2 +y^2+(1+\gamma ) z^2>}}= {\sqrt{1+\frac{2}{\gamma}}}~a \approx \left(1+\frac{1}{\gamma}\right)a .
    \label{12}
\end{equation}
A direct calculation of $R$ in (\ref{11}) using $<x>,<y>,<z>$ derived above, we find, for choice (I) 
\begin{equation}
    R_I=<x>_I=\frac{a}{2},
\end{equation}
and for choice (II) 
\begin{equation}
    R_{II}=<x>_{II}=\frac{a}{4\gamma}.
\end{equation}
Clearly, both choices satisfy the inequality (\ref{11}).

Finally we combine the three URs in (\ref{10}) to obtain
\begin{equation}
\sigma^2_{R}\equiv \left(\sigma_x^2 +\sigma_y^2 +\left(1+\frac{1}{\gamma}\right)\sigma_z^2\right)\sigma_{L_z}^2
\geq \frac{\hbar ^2}{4}\left<y+\frac{\alpha }{a}zx\right>^2 +
 \frac{\hbar ^2}{4}\left<x-\frac{\alpha }{a}zy\right>^2 +
 \left(1+\frac{1}{\gamma}\right) \frac{\hbar ^2\alpha^2}{4}\left<\frac{a}{\gamma}-\frac{\gamma}{2a}z^2\right>^2 .
     \label{10x}
\end{equation}
However, it is more natural to define a quantity $\tilde{\sigma}_{R}$, given by,
\begin{equation}
\tilde{\sigma}_{R}=\frac{{\sigma}_{R}}{R}
    \label{sd2}
\end{equation}
that turn out to be (to our $O(1/\gamma)$ approximation), for the two options,
\begin{equation}
 \tilde{\sigma}_{R(I)}\sigma_{L_z}\geq \frac{\hbar}{2},
    \label{y}
\end{equation}
\begin{equation}
 \tilde{\sigma}_{R(II)}\sigma_{L_z}\geq \frac{\hbar}{2}\left(1+\frac{q}{p}+\left(\frac{3q}{p}\right)^2\right) .
    \label{yy}
\end{equation}
Comparison between the above two URs suggests that, whereas for option (I), the normal UR is maintained for torus knots (in our approximate scheme), in option (II) the lower bound of the UR increases, suggesting that the motion of the quantum particle in a (torus) knot tend to increase the overall SDs and hence uncertainties.

\section{Conclusion and future outlook}
Let us summarize our framework and findings. We have studied Uncertainty Relations associated with a quantum particle on a torus knot. The motivation is two-fold: on one hand, the treatment of quantum mechanics of particles on a {\it{periodic path}} needs to be addressed very carefully. There are very few works in literature that study periodic motion in more complicated paths than a circular path. On the other hand, choosing a path embedded on a torus surface has many practical applications, apart from its theoretical interest; it is a nontrivial extension of the path on a circle, (we have followed the framework of the recent work \cite{on}), being in a higher dimensional configuration space and topological intricacies due to the presence of  {\it{paths with a knot}}. 

Commutation relations between the operators in question generate the  URs. For a periodic system, only periodic degrees of freedom can be used. For circular motion, a Cartesian parameterisation in $X-Y$ is sufficient but for the torus knot motion, a Cartesian parameterisation in three-dimensional $X-Y-Z$ space is necessary. As a result more involved URs are generated, which, however, can be reduced to the circular case under suitable (thin torus) approximation.   

The periodic paths on a torus have two periodicities (characterized by the integers $p,q$) compared to paths on a circle with a single periodicity, with $p,q$  obeying certain constraints for a knotted path.  For explicit computation, we consider as an example a superposition of two eigenfunctions as a generic wave function and verify the validity of the URs. Interestingly, we show that, even in this simple example, restrictions on $p,q$ yield two distinct superpositions are possible, giving rise to inequivalent URs. In all cases, the new Standard Deviations and URs involve (torus and knot parameter) corrections over the conventional values.

An interesting fact emerges that in the case of the SDs and URs, the local geometry of the knots plays the decisive role and not their topological properties. As an example note that the $(p,q)$ and $(q,p)$ torus knots are topologically equivalent but with different local geometries and our results are not invariant under the interchange of $p,q$.

The generalized results revealed here can be tested in experimental setups involving particle motion in fibres along knotted paths. In a theoretical context, it will be interesting to extend the present work in the case of fuzzy or non-commutative torus knots.
\section{Acknowledgement} It is indeed a pleasure to thank Professor Naohisa Ogawa for numerous helpful discussions. We thank the Referee for the constructive comments.

\end{document}